\documentclass{iaus}
\usepackage{graphicx}
\usepackage{natbib}

\title[S266.~~OC Characterization via Cross-Correlation] 
{Open Cluster Characterization via \\ Cross-Correlation with Spectral Library}

\author[Maia et al.] 
{Francisco\,F.\,S.\,Maia$^1$, Jo\~ao\,F.\,C.\,Santos\,Jr.$^1$,
  Wagner\,J.\,B.\,Corradi$^1$ and Andres E. Piatti$^2$}

\affiliation{$^1$Universidade Federal de Minas Gerais, ICEx - Dept. de F\'\i
  sica, Av. Ant\^onio Carlos, 6627, Belo Horizonte 31270-901, MG, Brazil \\
email: {\tt [ffsmaia,jfcsj,wag]@ufmg.br}
 \\[\affilskip]
$^2$Instituto de Astronom\'{\i}a y F\'{\i}sica del Espacio, CC67, Suc. 28,
 1428 Buenos Aires, Argentina \\ email: {\tt andres@iafe.uba.ar}}

\pubyear{2009}
\volume{266}  
\pagerange{119--126}
\jname{Star Clusters - Galactic Building Blocks throughout Time and Space}
\editors{Richard de Grijs \& Jacques R.~D. L\'epine, eds.}
\begin{document}

\maketitle

\begin{abstract}
We present a characterization method based on spectral cross-correlation to obtain the 
physical parameters of the controversial stellar aggregate ESO442-SC04. The data used 
was obtained with GMOS at Gemini-South telescope including 17 stars in the central 
region of the object and 6 standard-stars. {\sc fxcor} was used in an iterative process to 
obtain self-consistent radial velocities for the standard-stars and averaged radial 
velocities for the science spectra. Spectral types, effective temperature, suface gravity 
and metallicities parameters were determined using {\sc fxcor} to correlate cluster 
spectra with ELODIE spectral library and selecting the best correlation matches using the 
Tonry and Davis Ratio (TDR). Analysis of the results suggests that the stars in 
ESO442-SC04 are not bound and therefore they do not constitute a physical system.

\end{abstract}

\firstsection 
\firstsection

\section*{Introduction}

Open clusters are destroyed over time by the action of external forces (Galactic tidal 
field, collisions with molecular clouds) and as a consequence of their own dynamical and 
stellar evolution. Studies on 
the subject indicate that most open clusters dissolve in 0.5-2.5 Gyr \citep{Portegies01} 
and that finding clusters in the state of dissolution should be common.
However, discriminating such objects from a random over-density of field stars 
is a difficult task, and many authors have been
developing techniques to study and properly classify them \citep[e.g.,][]{Pavani07}.

ESO442-SC04 was identified by \citet{Bica01} as a Possible Open Cluster Remnant  and later was classified by \citet{Carraro05} as an asterism. We have applied an
iterative method to characterize this object 
and derive the physical parameters of its stars. 
\firstsection

\section*{Data}

The Gemini Multi-Object Spectrograph (GMOS - Gemini-South) was
used to collect spectra of 36 selected stars on a $5^\prime \times 5^\prime$
region centered in ESO442-SC04 and 6 standard-stars from \citet{Nordstrom}.
The collected spectra cover the spectral range $3875\,\mathrm{\AA} -
5300\,\mathrm{\AA}$ with resolution $R\approx4000$ and signal-to-noise in the range $5-50$.
Table \ref{t:std} lists the observed standard stars and their properties as found in the 
literature. 

\begin{table}[htb]
\centering \scriptsize
\caption{Literature data for the standard-stars. }
\begin{tabular}{|c|c|c|c|c|}
\hline
Object & ST & $ V_r $ (km/s) & [Fe/H] ($\pm$ 0.12) & T$_{eff}$ ($\pm$ 110) (K)  \\ \hline
HD104471  & G0V	 & -7.2 $\pm$ 0.1 & 0.00  & 5984 \\
HD104982  & G2V	 & 10.5 $\pm$ 0.1 & -0.40 &  5610 \\
HD105004  & F8VI & 121.6$\pm$ 0.3 & -0.79 & 5821 \\
HD107122  & F1V	 & 16.2 $\pm$ 3.3 & -0.42 &  6576 \\
HD111433  & F3IV & 4.0  $\pm$ 0.6 & 0.25  & 6471 \\
CD-289374 & ---	 & 30.4 $\pm$ 0.2 & -1.18 & 4830 \\ \hline
\end{tabular}
\label{t:std}
\end{table}

We adopted ELODIE.3.1 stellar spectral library \citep{Prugniel01} as template spectra in 
the cross-correlation technique to determine the physical parameters of the targets.
The library includes 1962 spectra of 1388 stars 
providing a large coverage of stellar atmospheric parameters $\mathrm{T_{eff}}$, 
$\log\,g$ and $[\mathrm{Fe/H}]$. 
\firstsection

\section*{Radial velocity determination}

Initially {\sc fxcor} task was applied on the standard-stars spectra using the values given 
by Table \ref{t:std} as a first guess to their radial velocity. A self-consistent set of solutions 
was then obtained by making corrections to the velocities and redoing the correlation in 
order to find consistent values for the whole set.
Table \ref{t:velrad}  shows the results of the correlations and the averaged radial velocities obtained.

\begin{table}[htb]
\centering \scriptsize
\caption{Calculated radial velocities for the standard stars. Boldfaced values represent 
the velocities adopted for each template in the correlation.}
\begin{tabular}{|c|c|c|c|c|c|c|}
\hline
{\tiny star/template} & HD104471 & HD104982 & HD105004 & HD107122 & HD111433 & CD-289374 \\
\hline
HD104471 &{\bf -126.3$\pm$11.5}&-7.5$\pm$5.2 &118.0$\pm$15.1&16.7$\pm$25.9&-24.6$\pm$11.4&30.4$\pm$5.7  \\
HD104982 &-127.5$\pm$5.2 &{\bf -8.7$\pm$5.7 }&114.0$\pm$13.4&9.4 $\pm$24.7&-28.8$\pm$13.0&32.9$\pm$11.2 \\
HD105004 &-122.7$\pm$15.8&-1.1$\pm$13.4&{\bf 121.6$\pm$0.3 }&19.6$\pm$27.9&-20.6$\pm$18.0&37.5$\pm$16.6 \\
HD107122 &-126.8$\pm$25.9&-1.9$\pm$24.7&118.2$\pm$27.9&{\bf 16.2$\pm$3.3 }&-24.6$\pm$18.4&36.0$\pm$28.9 \\
HD111433 &-126.3$\pm$11.5&-4.5$\pm$13.0&117.6$\pm$18.0&16.2$\pm$18.4&{\bf -24.6$\pm$11.4}&33.7$\pm$17.8 \\
CD-289374&-128.8$\pm$11.2&-8.7$\pm$5.2 &114.5$\pm$16.6&10.6$\pm$3.3 &-27.9$\pm$17.8&{\bf 30.4$\pm$0.2 } \\
\hline
$\overline{V_{r}}$&-126.4$\pm$6.1&-5.4$\pm$5.4&117.3$\pm$7.0&14.8$\pm$9.5&-25.2$\pm$6.3&33.5$\pm$6.6 \\
\hline
\end{tabular}
\label{t:velrad}
\end{table}

Then {\sc fxcor} was run on the science spectra using the 6 standard stars spectra as 
templates. The radial velocities found from each template were averaged and the 
dispersion calculated for all science spectra. Table \ref{t:results}  shows the used spectra, 
their corresponding S/N and the radial velocities obtained.
\firstsection

\section*{Cross-correlation with spectral library}

\begin{figure}[htb]
\centering
\includegraphics[width=0.75\linewidth]{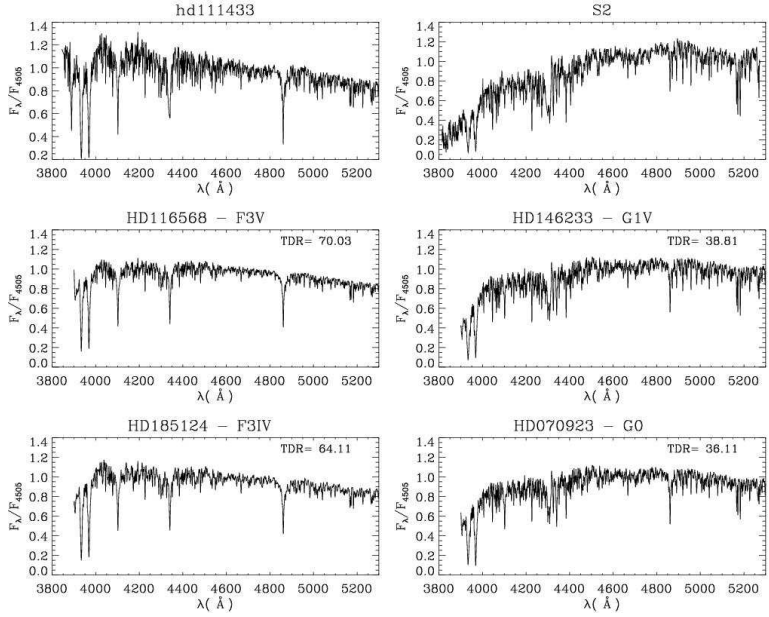}
\caption{Best correlation templates for standard-star HD\,111443 (left) and science 
spectra S2 (right). Name of the template and TDR value are shown.}
\label{f:tdr}
\end{figure}

We used ELODIE library spectra as templates to the standard-stars and the 
task {\sc fxcor} to determine, through the sharpness of the correlation function peak \citep[TDR,][]{Tonry}, the most similar templates to each star. Fig. \ref{f:tdr} 
shows the two best matching spectra found by correlation for the standard-star 
HD\,111443 and for one of our science-spectra. Their spectral types and TDR correlation 
parameter are also shown.

Spectral types were determined by collecting the ten most similar templates and 
summing over the TDR value of those with the same spectral type. 
The one with the highest TDR sum, was adopted as the final spectral type. Table \ref{t:spt} shows the determined spectral types for standard-stars in comparison with the 
values provided by SIMBAD.

\begin{table}[h]
\centering \scriptsize
\caption{Comparison of determined spectral types for standard stars.}
\begin{tabular}{|c|c|c|c|c|c|c|}
\hline
 & HD104471 & HD104982 & HD105004 & HD107122 & HD111433 & CD-289374 \\ \hline
Determined  & G0 & G2V   & F5V   & F3V    & F3V & G0  \\
SIMBAD      & G0V   & G2V      & F8IV  & F1V    & F3IV   & ---  \\
\hline
\end{tabular}
\label{t:spt}
\end{table}

\begin{figure}[htb]
\centering
\includegraphics[width=0.75\linewidth]{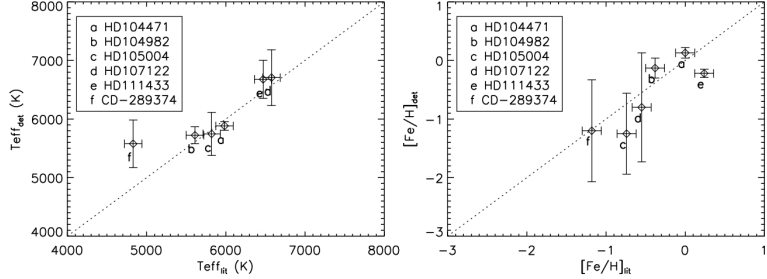}
\caption{Comparison between determined and literature parameters for the standard-
stars. Effective temperatures (left) and metallicities (right).}
\label{f:comp}
\end{figure}

Effective temperatures, surface gravity and metallicity were calculated as the 
averaged values listed for the chosen templates, weighted by the 
TDR correlation value of each template. The uncertainties adopted correspond to the 
weighted standard deviation of the derived average. Figure \ref{f:comp} shows the 
relation between the determined effective temperature and metallicity and the literature 
parameters.

We used the same method to 
determine the spectral type, effective temperature, surface gravity and metallicity 
of our science spectra, as shown in Table \ref{t:results}.

\firstsection \firstsection
\section*{Results}

By employing cross-correlation techniques and using ELODIE's spectral library we were 
able to determine heliocentric radial velocities, spectral types, effective temperatures,
surface gravity and metallicities for 17 stars in the inner region of stellar object 
ESO442-SC04. These results are summarized in Table \ref{t:results}.

\begin{table}[htb]
\centering \scriptsize
\caption{Radial velocity, spectral type, effective temperature, surface gravity and metallicity determined parameters for the science spectra.}
\begin{tabular}{|c|c|c|c|c|c|c|}
\hline
Spectrum & S/N &$\overline{{V_r}}$ ($\sigma_{V_r}$) (km/s) & Sp. Type & 
T$_{eff}$ ($\sigma_{T_{eff}}$) (K) & log(g) ($\sigma_{log(g)}$) & 
[Fe/H] ($\sigma_{[Fe/H]}$) \\ \hline
S2  & 15 & 75.8  (7.2 ) & G5V   & 5676 (117) & 4.20 (0.21) & -0.01 (0.09) \\
S6  & 40 & 9.5   (5.2 ) & F5V   & 6347 (348) & 4.24 (0.12) & -0.59 (0.27) \\
S9  & 20 & 25.1  (4.7 ) & F5V   & 6309 (416) & 4.24 (0.08) & -0.52 (0.29) \\
S10 & 10 & 93.6  (10.3) & K2V   & 5213 ( 61) & 4.48 (0.10) & -0.03 (0.14) \\
S12 & 20 & 227.2 (12.7) & G0    & 5380 (424) & 3.14 (0.88) & -1.60 (0.66) \\
S14 & 25 & 110.0 (10.9) & G0    & 5673 (324) & 3.75 (0.72) & -1.79 (0.57) \\
S15 & 30 & 78.5  (7.5 ) & G5V   & 5606 (115) & 4.33 (0.11) &  0.02 (0.08) \\
S19 & 50 & 68.0  (8.2 ) & F3V   & 6834 (261) & 4.28 (0.08) & -0.41 (0.29) \\
S20 & 25 & 74.4  (6.9 ) & G2V   & 5740 ( 63) & 4.17 (0.18) &  0.03 (0.13) \\
S21 & 30 & 31.3  (5.6 ) & F7IV  & 5910 (460) & 3.89 (0.75) & -0.99 (0.81) \\
S23 & 35 & 95.5  (8.5 ) & G0V   & 5765 (261) & 3.99 (0.56) & -0.92 (0.67) \\
S25 & 6  & 145.9 (17.1) & F2IV  & 6480 (439) & 4.14 (0.19) & -0.19 (0.16) \\
S28 & 30 & 15.0  (5.4 ) & F8V   & 6005 (306) & 4.19 (0.19) & -0.57 (0.26) \\
S29 & 22 & 10.7  (7.1 ) & G8III & 4877 ( 87) & 2.76 (0.16) & -0.03 (0.15) \\
S30 & 12 & 40.0  (6.6 ) & G0III & 5721 ( 69) & 4.07 (0.18) &  0.16 (0.07) \\
S32 & 10 & 47.0  (6.7 ) & G8III & 5011 (212) & 2.76 (0.42) & -0.03 (0.13) \\ 
S33 & 12 & 142.4 (12.9) & G8V   & 5633 (117) & 4.31 (0.18) & -0.01 (0.06) \\ \hline
\end{tabular}
\label{t:results}
\end{table}

\begin{figure}[htb]
\centering
\includegraphics[width=0.95\linewidth]{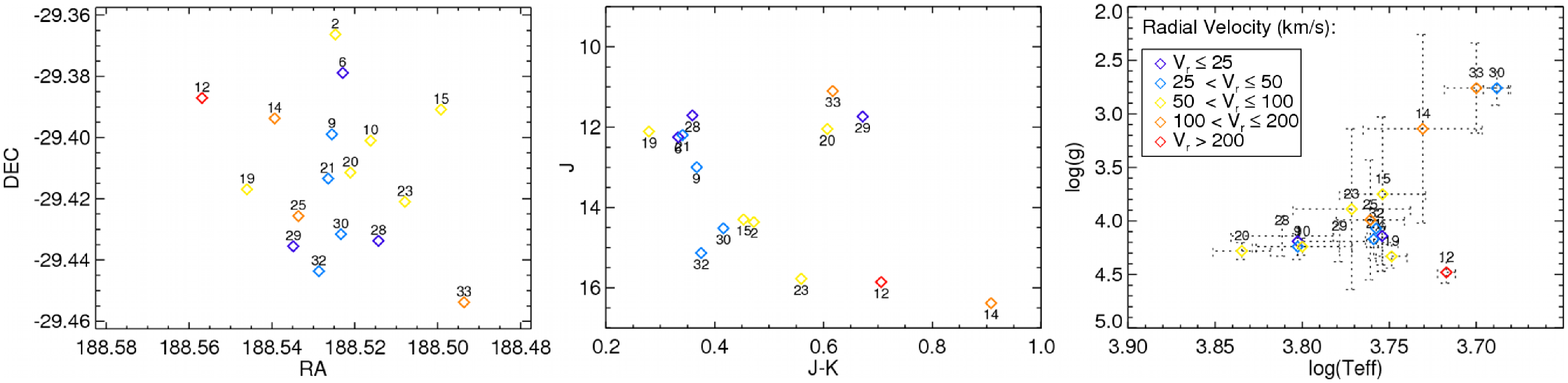}
\caption{Positional chart (left), color-magnitude diagram (middle) and surface 
gravity-temperature diagram (right) for the studied spectra. Stars are labeled by 
their internal identification shown in Table \ref{t:results} and color-coded according to 
their radial velocities (right legend).}
\label{f:diagrams}
\end{figure}

2MASS data was used to build a colour-magnitude diagram and
a spatial diagram showing the relative positions of the stars in the sky, along with the 
radial velocity, surface gravity and effective temperature data obtained for the targets. 
The objects in those diagrams are labeled according their internal identification 
and coded with colors representing groups with a range of radial velocities. Such 
diagrams are shown in Fig. \ref{f:diagrams}.

\firstsection

\section*{Conclusion}

We devised a spectral cross-correlation method to determine kinematical and 
astrophysical parameters for stars in the 
central region of ESO442-SC04.
Radial velocities show a dispersion greater than 50 km/s and the metallicities show 
a dispersion greater than 0.6 dex, suggesting that few, if any, of the studied stars are 
physically bound. 
Although the stars distribution in the CMD (Fig. \ref{f:diagrams}, middle) suggests that 
they follow evolutionary sequences typical of old clusters, our results suggest that 
ESO442-SC04 does not constitute a physical system.

\end{document}